\documentclass[a4paper,11pt]{article}
\pdfoutput=1 % if your are submitting a pdflatex (i.e. if you have
\usepackage{jinstpub} % for details on the use of the package, please
\usepackage{xcolor}
\usepackage{siunitx}
\usepackage[inline]{enumitem}
\usepackage{textgreek}
\usepackage{xfrac}
\usepackage{systeme}
\usepackage{wrapfig}

\graphicspath{{./images/}}
 
\title{Validating the ASCOT modelling of NBI fast ions in Wendelstein 7-X stellarator}

\author[a,1]{S. \"{A}k\"{a}slompolo,\note{Corresponding author.}}

\author[a]{P. Drewelow}
\author[b]{Y. Gao}
\author[a,2]{A. Ali\note{Current affiliation: Max Planck Institute of Molecular Physiology, Germany}}
\author[a]{C. Biedermann}
\author[a]{S. Bozhenkov}
\author[a]{ C.P. Dhard}
\author[a]{M. Endler}
\author[a]{J. Fellinger}
\author[a]{O.P. Ford}
\author[a]{B. Geiger}
\author[a]{J. Geiger}
\author[d]{N. den Harder}
\author[a]{D. Hartmann}
\author[a]{D. Hathiramani}
\author[g]{M. Isobe}
\author[a]{M. Jakubowski}
\author[h]{Y. Kazakov}
\author[a]{C. Killer    }
\author[j]{S. Lazerson}
\author[d]{M. Mayer}
\author[a]{P. McNeely}
\author[a]{D. Naujoks}
\author[a]{T.W.C. Neelis}
\author[c]{J. Kontula}
\author[c]{T. Kurki-Suonio}  
\author[a]{H. Niemann}
\author[g]{K. Ogawa}
\author[e]{F. Pisano}
\author[a]{P. Zs. Poloskei}
\author[a]{A. Puig Sitjes}
\author[a]{K. Rahbarnia}
\author[a]{N. Rust}
\author[i]{J. C. Schmitt} % Auburn University
\author[f]{M. Sleczka}
\author[a]{L. Vano}
\author[a]{A. van Vuuren}
\author[k]{G. Wurden}
\author[a]{R.C. Wolf}

\affiliation[a]{Max-Planck-Institut f\"{u}r Plasmaphysik, Teilinstitut Greifswald, Wendelsteinstra\ss{}e 1, 17491 Greifswald Germany}
\affiliation[b]{Forschungzentrum J\"{u}lich, IEK-4, 52425 J\"{u}lich, Germany}
\affiliation[c]{Aalto University, 00076 Aalto, Finland}
\affiliation[d]{Max-Planck-Institut f\"{u}r Plasmaphysik, Boltzmannstra\ss{}e 2, 85748 Garching, Germany}
\affiliation[e]{University of Cagliari, Via Universit\`{a}, 40 - 09124 Cagliari, Italy}
\affiliation[f]{University of Szczecin, al. Papie\.{z}a Jana Paw\l{}a II 22a, 70-453 Szczecin, Poland}
\affiliation[g]{National Institute for Fusion Science, 322-6, Oroshi-cho, Toki-city, Gifu, Japan}
\affiliation[h]{Laboratory for Plasma Physics, LPP-ERM/KMS, Brussels, Belgium}
\affiliation[i]{Auburn University, Auburn, AL 36849, USA}
\affiliation[j]{Princeton Plasma Physics Laboratory, Princeton, NJ, USA}
\affiliation[k]{LANL, Los Alamos, NM  USA}

\emailAdd{simppa.akaslompolo@alumni.aalto.fi}

\abstract{
The first fast ion experiments in Wendelstein 7-X were performed in 2018. They are one of the first steps in demonstrating the optimised fast ion confinement of the stellarator. The fast ions were produced with a neutral beam injection (NBI) system 
and detected with infrared cameras (IR), a fast ion loss detector (FILD), fast ion charge exchange spectroscopy (FIDA), and post-mortem analysis of plasma facing components. 

The fast ion distribution function in the plasma and at the wall is being modelled with the ASCOT suite of codes. They calculate the ionisation of the injected neutrals and the consecutive slowing down process of the fast ions.
The primary output of the code is the multidimensional fast ion distribution function within the plasma and the distribution of particle hit locations and velocities on the wall. 
Synthetic measurements based on ASCOT output are compared to experimental results to assess the validity of the modelling.

This contribution presents an overview of the various fast ion measurements in 2018 and the current modelling status.
The validation and data-analysis is on-going, but the wall load IR modelling already yield results that match with the experiments.

}

\keywords{Plasma diagnostics - charged-particle spectroscopy; Plasma diagnostics - interferometry, spectroscopy and imaging; Simulation methods and programs}

\collaboration[c]{and the W7-X Team}

\proceeding{3$^{\text{rd}}$ European Conference on Plasma Diagnostics
\\
6$^{\text{th}}$ to 9$^{\text{th}}$ of May 2019,\\ 
Lisbon, Portugal 
}

\begin{document}
\maketitle
\flushbottom

\section{Introduction}
\label{sec:intro}

The Wendelstein 7-X stellarator  (W7-X) is a toroidal magnetic confinement fusion device.
In the stellarator, the toroidal magnetic field is produced predominantly by external 3D shaped coils, which removes the approximate toroidal symmetry present in the more main-stream tokamak-like devices.
The lack of symmetry results in challenges in the confinement of particles, in particular of the weakly-collisional fast ions.
Any fusion reactor with a burning plasma must confine the fusion product $\upalpha$-particles, and thus good fast ion confinement is vital for future reactors.
Thus, Wendelstein 7-X is optimised to tackle the fast ion confinement challenge at high plasma pressure.
Demonstration of this optimisation is one of the high-level goals of the project.

W7-X saw its first fast ions in Summer 2018, when first of the two NBI boxes was commissioned with two NBI sources.
Both sources injected hydrogen with up to \SI{55}{\kilo\electronvolt} energy for up to \SI{5}{\second} and delivered up to \SI{1.7}{\mega\watt} per source through the port.
A number of diagnostics saw a clear signal due to the fast ions: infra-red cameras (IR), dedicated fast ion loss detector (FILD) and charge exchange recombination spectroscopy (CXRS) spectrometers.
Fast ion effects in the coils measuring the plasma current and the diamagnetic energy are more difficult to analyse.

Validated fast ion modelling tools are needed to prepare future fast ion diagnostics and to confirm the improved fast ion confinement of W7-X.
This contribution gives an overview of the work on validating the ASCOT suite of fast ion modelling codes against measurements in the 2018 W7-X fast ion experiments. First the codes are introduced in section~\ref{sec:ascot}, then validation of the fast ion distribution function is discussed in section~\ref{sec:confined}. 
The code predicts localised NBI orbit loss hot spots and thus, section~\ref{sec:armor} describes the related effects to the plasma facing components.
Comparison against measured and synthetic FILD signals are given in section~\ref{sec:FILD}.
The work is still on-going, which is discussed in the last section~\ref{sec:summary}.

%The fast ions carry a special significance due to the well known challenges of fast ion confinement in classical stellarators.

%\subsection{W7-X \& NBI}

%\section{The lost fast ions}

\section{The ASCOT suite of codes}
\label{sec:ascot}
The ASCOT suite of codes is mainly used to model fast ions in tokamaks and stellarators.
The fast ions are usually NBI ions, fusion $\upalpha$-particles or, recently also, ion-cyclotron resonance heated ions.
In the current context of W7-X, the two most relevant programs are Beamlet Based NBI (BBNBI)~\cite{Asunta14_BBNBI_reference} and ASCOT4~\cite{ascot4ref}, that calculate the NBI ionisation and fast ion slowing down distribution (the Fokker-Planck equation), respectively.

%\subsection{BBNBI}
The W7-X NBI model within BBNBI mainly consists of the geometry of the 774 apertures in the NBI acceleration grid and of the detailed beam duct and dump shapes. The geometry is validated in \cite{emptyTorusW7X}. 
BBNBI uses Monte-Carlo methods to generate ionisation locations of the injected neutrals after they have passed through the beam duct.
The ionisation locations are the initial positions for the markers followed by ASCOT4 as either guiding centres or with full gyro-motion resolved.
Each marker represents multiple real particles, as denoted by the weight of the marker. 
ASCOT4 evaluates the local plasma temperature, density, magnetic and electric field values to follow the phase space paths of the markers, which represent the fast ion distribution function within ASCOT.
The marker following is stopped if the particle slows down to the local thermal energy or hits the wall.
%\subsection{ASCOT4}

The ASCOT4 code follows the markers all the way to the wall and collects the final states of the markers upon wall-contact.
The detailed W7-X wall used in the simulations consists of several million triangles with various sizes ranging from \SI{0.1}{\milli\meter\squared} to \SI{0.2}{\meter\squared} depending on the required level of detail.
A typical simulation for wall load analysis uses $10^7$ markers, from which approximately \SIrange{10}{15}{\percent} reach the wall.
The bulk of wall hits occur after diffusion through phase space and there is a wide distribution of energies from NBI injection energy to the minimum followed energy.
Post-processing the wall hits allows calculation of wall power loads as well as certain synthetic diagnostics signals.

In this work, the magnetic fields and plasma density and temperature profiles were acquired as equilibrium reconstructions using the V3FIT code~\cite{V3FIT_REF_2009}. The plasma consists of hydrogen with carbon as impurity. 
The carbon content was calculated by assuming the effective charge $Z_\textrm{eff}=\sum_in_iZ_i^2/\sum_in_iZ_i=1.5$, where $n_i$ and $Z_i$ are the density and charge of ion species $i$.
The Neo-Classical radial electric field was calculated from V3FIT profiles using the NTSS code~\cite{NTSS_turkin_transport_code}.

%\subsection{Acquiring the inputs}
%\subsubsection{V3FIT}

%\subsubsection{NTSS}

%\section{NBI injection}
%\subsection{IR camera measurements of the shinethrough}
%\subsection{NBI density from BES}

\section{The confined fast ions}
\label{sec:confined}

ASCOT represents the fast ion distribution as an ensemble of markers.
As the code follows the phase-space paths of the markers, it collects multi-dimensional histograms of the density of these paths.
These histograms represent various moments of the fast ion distribution function.
Post-processing them, possibly with additional information, facilitates the calculation of various synthetic diagnostic signals.
The parallel to magnetic field part of the fast ion velocity distribution contributes to the (neutral beam) fast ion current drive (NBCD) while the perpendicular component contributes to the plasma diamagnetic energy. 
Both of these quantities are measured with magnetic pickup coils in W7-X and, in principle comparison between ASCOT and measurements is possible.
However, both signals are heavily entangled with the stronger contributions from thermal ions and electrons.

The contribution of NBCD to plasma current is effectively masked by the following three contributions: the electron shielding current, plasma inductance, and varying bootstrap current due to changing profiles.
ASCOT does predict that the NBCD should have a negative sign, which is confirmed by the experimental observation that the plasma current always starts to decrease when NBI is turned on. The ASCOT predicted steady state NBCD current has not been reached in the experiments due to the long timescales associated with current evolution compared to the limited NBI pulse length.

The diamagnetic energy of the fast ions should appear and dissappear in the diamagnetic measurements on the fast ion slowing down/confinement time scale (e.g. \SI{20}{\milli\second}) from the turning on or off the beams.
Nevertheless, no step matching in timescale and/or amplitude predicted by ASCOT was observed in the studied discharges.
The fast ion contribution is most likely embedded in the change due to changing profiles.
Extracting the fast ion contribution from the plasma current or diamagnetic energy would require careful transport or equilibrium reconstruction analysis, which has not yet been performed with the required precision.

The fast ion distribution function was also measured using the Doppler shifted line radiation from the fast ions neutralised in the NBI neutral cloud. 
The so-called FIDA measurement uses charge exchange recombination spectroscopy (CXRS) sightlines and spectrometers to measure the local fast ion velocity distribution projection on the line-of-sight.
The analysis of the spectra using the FIDASIM code~\cite{heidbrink2010code} is on-going.
FIDASIM reproduces the experimental spectral hydrogen Balmer-$\upalpha$ spectrum shape (figure~11 of~\cite{PerformanceW7XplasmasDuringFirstDivertorOP}.
However, the magnitude of the signal, especially the fast ion contribution, is sensitive to the input parameters and plasma profiles, which are still under investigation. 
Thus, a quantitative fast ion match is not yet achieved.
%{\color{red}The fast ion contribution is visible in the spectrum, however the spectrum is }
%\subsection{Diamagnetic energy and toroidal current}
%\subsection{FIDASIM}

\section{Armouring against orbit losses and forensic analysis of a melting event}
\label{sec:armor}
The ASCOT simulations predicted high heat loads on certain plasma facing components (PFCs) already before the first NBI experiments. 
Heat loads to certain immersion tubes with sensitive vacuum windows were proactively armoured~\cite{armoringW7XimmersionTubes} to prevent leaks due to thermal stresses.
Infra red (IR) monitoring during experiments indicated heat loads in excess of \SI{1.5}{\mega\watt\per\meter\squared}~\cite{armoringW7XimmersionTubes}.
The armouring expanded the safe operation space of the NBI by allowing longer NBI pulses without endangering the machine plasma facing components.  In fact the NBI pulse length was ultimately extended to the NBI injector design value of 5s using a staged approach.

In contrast, there was an un-planned melting event due to NBI orbit losses. 
This involved one of the 44 wafer holders that were installed onto the stainless steel wall protection panels in one toroidal period of the machine.
These sugar-cube ($\sim$\SI{1}{\centi\meter\cubed}) sized sheet-metal components protrude from the smooth steel panel by roughly a centimetre. 
This is enough to skim the flux of NBI orbit loss particles that otherwise spread the load on the surface of the panel. 
Hence, the power was now deposited to the small cross-sectional area at the leading edge of the holder, which partly melted during the campaign~\cite{Naujoks2019proceedingsOfBadHonnef}.

%Figure~\ref{fig:meltedHolder} illustrates ASCOT simulation results how heat loads to the 44 sample holders. 
A number of ASCOT simulation were performed to assess the situation resulting high heat loads to the sample holders. % how heat loads to the 44 sample holders.
Remarkably, only the melted holder receives a significant power load in any of the simulations, approximately \SI{200}{W}, which is roughly the correct power needed to melt the holder in one second.
The load was present only in the simulations of the so-called standard magnetic configuration~\cite{VacuumMagneticConfigurationsW7X}.
The exact experiment(s) where the melt occurred have not been identified.

%\begin{figure}
%    \centering
%    \includegraphics[width=\textwidth]{images/sampleSummary_3d_run3467817_20180918_040.png}
%    \caption{{\color{blue}Melted sample holder related ASCOT simulation results. Say that this is one of a few magnetic confs. Say that we don't know the exact melting event.}}
%    \label{fig:meltedHolder}
%\end{figure}

\section{IR camera measurements of orbit losses}
\label{sec:IR}
The W7-X stellarator has a wide-angle infra-red camera system~\cite{IRimagingSys4wallProtectionInW7X} for monitoring the plasma facing components.
Several NBI orbit loss hot-spots are visible in the IR images, (see figure~10 in~\cite{PerformanceW7XplasmasDuringFirstDivertorOP}).
%\begin{itemize}
%    \item Configuration dependency
%    \item Quantitative agreement in figure~\ref{fig:orbitLossesQuantitative}
%\end{itemize}
%See \cite{WolfAPS2018proceedings} figure XX how well
%\begin{figure}
\begin{wrapfigure}[26]{r}{0.5\textwidth}%
%\centering
\includegraphics[trim=0cm 0cm 0cm 11mm,clip,width=0.5\textwidth]{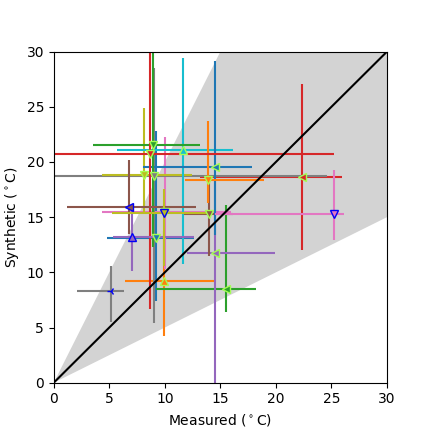}%
\caption{Temperature change in hot spots due to NBI orbit losses:  ASCOT modelling vs IR camera measurements. The measurement shows the temperature difference between an IR camera frame before and after an NBI pulse. The synthetic modelling includes only the NBI contribution, while the measurement may include changes due to thermal particle loads. The shown figure is created by analysing hand-picked hot-spots from pairs of synthetic and measured IR images. Only spots clearly visible in both images and not overlapping obviously thermal hot-spots were included in the analysis. The median, maximum and minimum values of the nine pixels nearest to the spot are used as the values and error bars in the graph. The grey shading indicates the area within a factor of two.}
    \label{fig:orbitLossesQuantitative}
\end{wrapfigure}
%\end{figure}
ASCOT predicts that the 
relative strength of these hot-spots depend on the magnetic field configurations, which was experimentally confirmed~\cite{PerformanceW7XplasmasDuringFirstDivertorOP}.
The stainless steel surfaces were generally too reflective for measuring the heat load due to NBI orbit losses.
However, multiple hot-spots on carbon PFCs were clearly identifiable by comparing the surface temperatures just before and after an NBI pulse.
Figure~\ref{fig:orbitLossesQuantitative} presents a comparison of predicted and measured hot-spot temperature changes.
The presented data-set provides an unprecedented experimental validation of fast ion orbit loss calculation in fully 3D magnetic field.
The synthetic IR camera model is validated in \cite{emptyTorusW7X}.

\section{Dedicated fast ion orbit loss detector}
\label{sec:FILD}

A dedicated fast ion orbit-loss measurement was performed with a fast ion loss detector (FILD)~\cite{fildOgawaECPD2019}.
The probe head was installed to the multi-purpose manipulator and consists of a graphite and molybdenum cover protecting 8 Faraday-cup detectors. 
The fast ions are admitted into the probe via a small aperture and then through a collimator. 
Thus, the different detectors sample particles with different gyroradii and pitches (i.e. ratio of parallel velocity to total velocity, $\xi=v_\parallel/v$).

The ASCOT simulations of the FILD probe signal follow a three-step method. 
First, a detailed CAD model of the probe is included in an ASCOT simulation and "near-hit" markers almost hitting the aperture are used as input to the next phase.
Next, the aperture is populated with a large number of markers uniformly distributed in phase space. 
Their trajectories are followed inside the probe head to detect the particles hitting each detector.
Now each "near hit" marker weight is multiplied by $p\cdot A_\text{A}/A_\text{nh}$, where  $p$ is the probability of a particle with similar velocity entering the aperture has to hit each detector. Finally, $A_\text{A}$ and $A_\text{nh}$ are the areas of the aperture and the surface where near hits are accepted, respectively.
The final step is to make a histogram of the times it took near-hit markers to reach the probe after ionisation and make a convolution-like combination with a rectangle function representing the NBI heating pulse.
The final result is then a time-dependent synthetic current signal for each detector.

\begin{figure}
    \centering
    \includegraphics[trim=2cm 61.8mm 18mm 149mm, clip,width=0.9\textwidth]{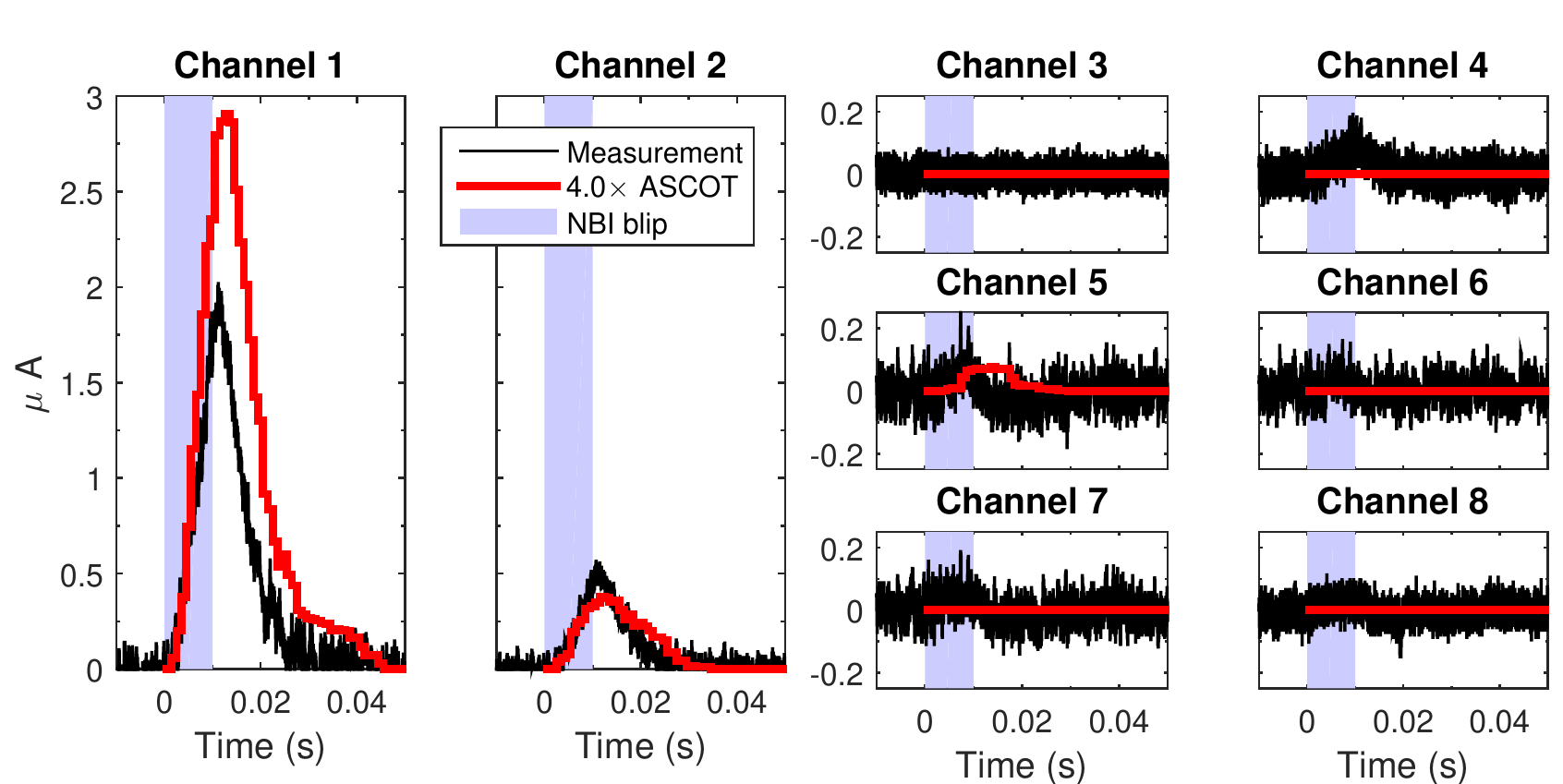} 
    \caption{Synthetic and measured FILD signal averaged over four NBI blips. 
    The scale for channels 1--2 is much larger than for the rest, which implies the signal in channels 3--8 should be considered noise.
    The channels 1--4 detect smaller lost particle energies than channels 5--8~\cite{fildOgawaECPD2019}. The particle pitches increase within channel sequences 1--4 and 5--8~\cite{fildOgawaECPD2019}. The data is from program 20180918.45 @4.7s. A DC component is removed from measured currents and ASCOT results have been multiplied by 4. }
    \label{fig:FILD}
\end{figure}

Figure~\ref{fig:FILD} shows a comparison of the synthetic FILD signals and averages over four \SI{10}{\milli\second} NBI blips.
The time trace shape of the signal corresponds to the experimentally measured signal. 
More specifically, the characteristic decay time of the signal is well matched, which indicates  that the diffusion process from ionisation to the probe is modelled essentially correctly.
The relative amplitudes of the different channels match well, but the total amplitude of the signal is underestimated by a factor of four. 
The prime candidate for explaining the missing signal is related to the magnetic field details near the probe: the effective distance from the plasma may be incorrect e.g. due to error fields, problems in the equilibrium or calibration errors. 

The flux of fast ions is a function of distance from the  plasma. 
In the modelled program, measurements~\cite{fildOgawaECPD2019} were conducted for three FILD radial locations ($R$=6.141, 6.127,  \SI{6.112}{\meter}), with each step moving the probe closer to the plasma.  Experimentally, the fast ion signal has a maximum at the middle $R$. ($I_\text{ch.1}^\text{meas}$= 1.6, 2.4, \SI{1.9}{\micro\ampere}). 
However, the modelling shows a minimum in the middle $R$. ($I_\text{ch.1}^\text{model}$= 0.7, 0.6, \SI{0.7}{\micro\ampere}). 
One possible explanation for the discrepancy comes from inaccuracies in the used magnetic field, that may lead to excessive effective radial distance from the plasma. 
A detailed future study is needed to completely interpret the measurements.

% 1.6 2.4 1.9
% 0.7 0.6 0.7

\section{Discussion and Summary}
\label{sec:summary}

This contribution gives an overview of the on-going first attempts in validating the orbit following Monte-Carlo code ASCOT against NBI fast ion measurements in the Wendelstein 7-X stellarator.
Both the fast ion distribution function as well as the wall loads are compared against measurements.
The most successful, quantitative, comparisons are performed utilising a fast ion loss detector (FILD) and infrared thermography.
Many plasma and NBI parameters of W7-X are still preliminary, while some are only assumptions based on similar machines.
Thus, the work presented can only be the first iteration of the on-going validation work.
Furthermore, certain relatively simple NBI measurements with plasma are not yet performed: shine-through power and NBI density from beam emission spectroscopy models should be compared to measurements.
Detailed analysis of the FIDA and FILD modelling results may be analysed in detail in separate contributions.

%\begin{itemize}
%    \item Future work: BES
%    \item The more details one checks, more facts become assumptions
%    \item FILD, FIDA analysis worth their own articles, perhaps also IR measurements
%    \item 
%\end{itemize}

 \acknowledgments
 This work was funded partly by Walter Ahlstr\"{o}m foundation. This work has been carried out within the framework of the EUROfusion Consortium and has received funding from the Euratom research and training programme 2014-2018 and 2019-2020 under grant agreement No 633053. The views and opinions expressed herein do not necessarily reflect those of the European Commission.

\bibliographystyle{unsrturl}
\bibliography{bibfile}
\end{document}